\documentclass{kluwer}
\usepackage[]{graphicx}

\begin{document}
\begin{article}
\begin{opening}

\title{Did Open Solar Magnetic Field Increase during the Last 100 Years:
A Reanalysis of Geomagnetic Activity}

\author{K. \surname{Mursula}}
\institute{Department of Physical Sciences, University of Oulu,
Finland; email: Kalevi.Mursula@oulu.fi}

\author{D. \surname{Martini}}

\institute{Department of Physical Sciences, University of Oulu,
Finland, and GGKI, SOPRON , Hungary }

\author{A. \surname{Karinen}}
\institute{Department of Physical Sciences, University of Oulu,
Finland}

\runningauthor{K. Mursula, D. Martini, and A. Karinen}
\runningtitle{Geomagnetic activity during the last 100 years}

%\date{(Received ?? November 2003; accepted ....)}

\begin{abstract}
Long-term geomagnetic activity presented by the aa index 
has been used to show that the heliospheric magnetic field has 
more than doubled during the last 100 years. 
However, serious concern has been raised on the long-term consistency 
of the aa index and on the centennial rise of the solar magnetic field. 
Here we reanalyze geomagnetic activity during the last 100 years by  
calculating the recently suggested IHV 
(Inter-Hour Variability) index as a measure of 
local geomagnetic activity for seven stations.
We find that local geomagnetic activity at all stations follows the same 
qualitative long-term pattern:
an increase from early 1900s to 1960, a dramatic dropout in 1960s
and a (mostly weaker) increase thereafter.
Moreover, at all stations, the activity at the end of the
20th century has a higher average level than at the beginning of the century.
This agrees with the result based on the aa index
that global geomagnetic activity, and thereby, the open solar magnetic field 
has indeed increased during the last 100 years.
However, quantitatively, the estimated
centennial increase varies greatly from one station to another.
We find that the relative increase is higher at the 
high-latitude stations and lower at the low and mid-latitude stations. 
These differences may indicate that the fraction of  
solar wind disturbances leading to only moderate geomagnetic activity
has increased during the studied time interval. 
We also show that the IHV index needs to be corrected for the long-term change
of the daily curve, and calculate the corrected IHV values.  
Most dramatically, we find the centennial increase in
global geomagnetic activity was considerably smaller,
only about one half of that depicted by the aa index.

\end{abstract}


\end{opening}

%____________________SECTION_1______________
\section{Introduction}

% TABLE I
\begin{table} 
%\tiny
\small
%\begin{table*} %  Star means there is no numbering of Table
%\begin{center}
%\begin{tabular}{|l|c|c|c|c|c|c|} \hline
\begin{tabular}{lrrrrrrr} \hline
Station & \multicolumn{1}{c}{IAGA} & \multicolumn{2}{c}{Geographic} & 
\multicolumn{2}{c}{Geomagnetic} &\multicolumn{1}{c}{Midnight} &\multicolumn{1}{c}{Data}\\
 Ê&\multicolumn{1}{c}{Code} & \multicolumn{1}{c}{Lat} & \multicolumn{1}{c}{Long} & \multicolumn{1}{c}{Lat} &
\multicolumn{1}{c}{Long} & \multicolumn{1}{c}{UT hour} &
\multicolumn{1}{c}{start} \\
\hline
Sodankyl{\"a} & SOD & 67.47 & 26.60 & 63.96 & 120.25 & 22 & 1914\\ %\hline
Sitka  &  SIT & 57.05 & 224.67  & 60.33 & 279.79 & 9 & 1902\\  %\hline
Eskdalemuir & ESK & 55.32 & 356.80 & 57.86 & 83.85 & 0 & 1911 \\ %\hline
Niemegk & NGK & 52.07 & 12.68 & 51.89 & 97.69 & 23 & 1901 \\  %\hline
Cheltenham & CLH & 38.73 & 283.16 & 49.14 & 353.71 & 5 & 1901 \\  %\hline
Fredericksburg & FRD & 38.20 & 282.63 & 48.59 & 353.11 & 5 & 1956 \\  %\hline
Tucson & TUC & 32.25 & 249.17 & 40.06 & 315.63 & 7 & 1909 \\  %\hline
Honolulu  & HON & 21.31 & 201.91 & 21.57 & 269.37 & 10 & 1902\\
\hline
\end{tabular}
\caption[]{ Information on stations used. Magnetic coordinates are calculated 
using the IGRF 2000 model.}
%\end{center}
\end{table}

One of the most interesting and important questions in solar-terrestrial physics 
is whether the magnetic activity of the Sun has indeed 
greatly increased during the last 100 years.  
A significant increase in solar
magnetic activity is indicated, e.g., by the well known fact  that the average
amplitude of sunspot cycles during the latter part of the  20th century is higher
than in the beginning.  
The increasing sunspot activity leads, according to a
simple model presented by  Solanki et al. (2000), to a corresponding long-term
increase in the total solar magnetic field, as well as in the open solar
magnetic field, i.e., in the heliospheric magnetic field 
(also called the interplanetary magnetic field).

Heliospheric  magnetic field is known to be 
the main modulator of cosmic rays and a significant 
cause of geomagnetic activity.
Lockwood et al. (1999) used the long-term geomagnetic aa index to show that the heliospheric 
magnetic field is now more than twice stronger than 100 years ago.
Recently, using long-term measurements of 
cosmogenic isotopes and a sophisticated chain of physical models,
it was shown (Usoskin et al., 2003) that the solar cycle 
averaged sunspot activity  since 1940s is higher than at any other time
during the last 1000 years.

Despite the versatility and uniformity of these important results, 
serious  concern has recently been raised on the centennial rise of solar
magnetic activity. 
E.g., the long-term consistency of the geomagnetic aa index on
which perhaps the most reliable evidence on centennial rise is based, has been
seriously questioned (Clilverd et al., 2002). 
Because of these problems with the aa index,
Svalgaard et al. (2003) introduced the IHV index as a more
straightforward and homogeneous measure of local long-term geomagnetic activity.
Using the data from the Cheltenham/Fredricksburg station pair, they 
found no evidence for an increase in the corresponding 
IHV index during the last 100 years. 

In this paper we calculate the IHV index for several geomagnetic stations 
in order to obtain a more global and reliable view on the
long-term development of global geomagnetic activity and in order to
conclude whether or not there is reliable evidence for a centennial 
increase of geomagnetic activity and, thereby, of the open solar magnetic field.
Moreover, we study in detail the daily variation of the magnetic field,
its long-term change and its effect on the IHV index.

%__________________Figure 1____________________________________________
%
\begin{figure}
\centerline{\includegraphics[width=24pc]{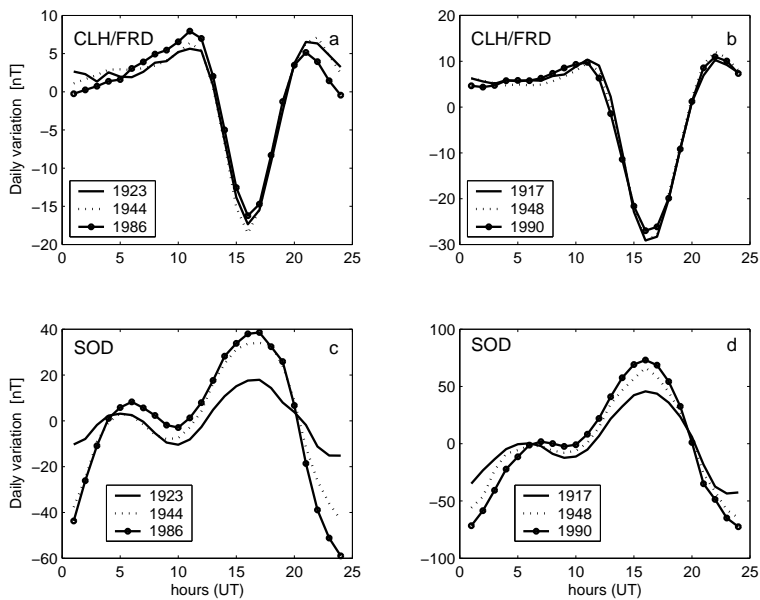}}
%\centerline{\includegraphics[width=18pc, angle=270]{fig1eps.eps}}
\caption{Daily curves for CLH/FRD and SOD stations for some 
sunspot minimum (a and c) and maximum (b and d) years.
Time is given in UT hours.}
\end{figure}
%
%_______________________________________________________________________
%

%____________________SECTION_2______________

\section{Data and daily variation}

We use here data from six stations and one station pair which have a long record
of  magnetic observations from early 1900s.
The codes, coordinates, local midnight UT hours, and start years of these
stations are depicted in Table I.  (The stations are listed in the order of
decreasing latitude). CLH and FRD form a station pair and is included here in
order to allow a comparison with the results by Svalgaard et al. (2004). 

Figure 1 depicts the yearly averaged daily variation curves of the H-component 
for CLH/FRD and SOD for some sunspot minimum and maximum years.
(The daily average was subtracted in each case in order to remove the
effect of the secular variation of the magnetic field).
One can see that while the daily curves at the two stations are very different
(even if shifted so as to coincide in LT time), they remain
fairly similar in form over the nearly 100 year time interval. 
We remind that the daily curve is, especially at low and mid-latitudes like
CLH/FRD,  mostly due to the dayside ionospheric $S_q$ (quiet day) current.

Figure 1 shows that although the form of the daily curve remains 
roughly the same each year, 
its range (difference between daily maximum and minimum) 
varies greatly, especially between the minimum and maximum sunspot years. 
Figure 2 depicts the yearly averaged range of the daily curve for a number of
stations over the whole time interval. 
The variation of the daily range with solar cycle is apparent.  
Moreover, one can
see in Figure 2 that the similarity between the daily range and sunspot number
curves (and the correlation between the corresponding values) is excellent for the
low and mid-latitude stations (Figure 2 c and d). 
This is in accordance with the fact that the $S_q$ variation
is mainly due to solar UV radiation, and thereby, is closely correlated with
sunspot numbers. 
However,  at high latitudes (Figure 2 a and b), where current systems 
other than the $S_q$ current start dominating the daily
curve, the daily range curves correlate better with geomagnetic activity  than sunspot numbers.

%__________________Figure 2____________________________________________
%
%
\begin{figure*}
      \begin{center}
      \resizebox{\hsize}{!}{\includegraphics{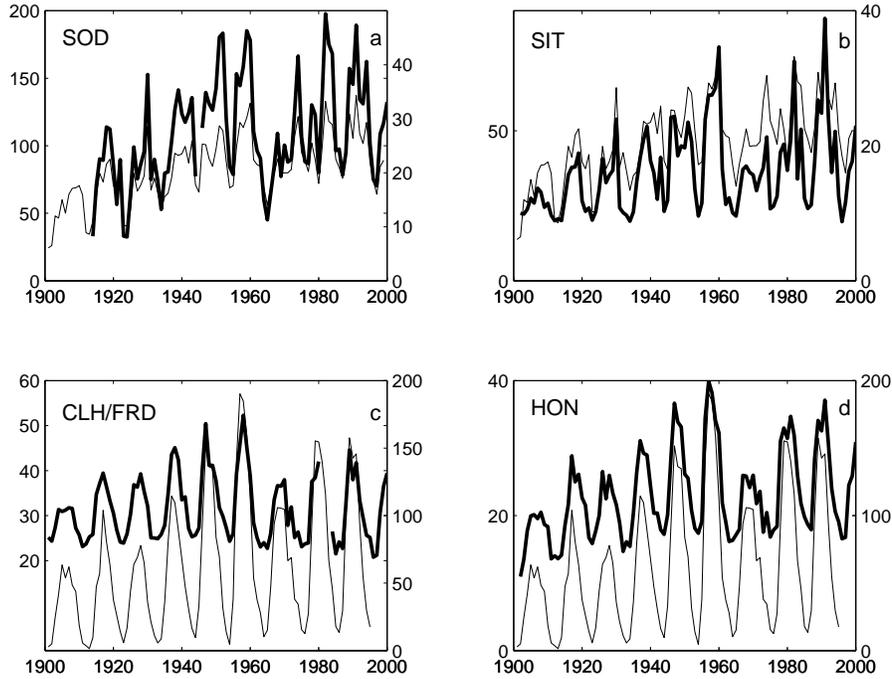}}
      \end{center}
      %\vskip -0.4cm
\caption{Yearly averages of the daily range at 4 stations (left axis in nT units, thick lines)
together with the aa index (panels a and b; right axis in nT units, thin lines) and sunspot
numbers (panels c and d; right axis, thin lines). }
\label{Daily_ampls}
%\vskip -0.3cm
\end{figure*}

%
%_______________________________________________________________________
%

%____________________SECTION_3______________

\section{Original IHV index}

Figure 1 shows that the CLH/FRD daily curve has a smooth section 
(almost a plateau) in the local pre-midnight hours from 00 to 06 UT 
prior to an increase at about 10 UT and a large depletion at local noon. 
Because of this smoothness (and because this time coincides with the 
geomagnetically active pre-midnight local sector) 
Svalgaard et al. (2004) used the CLH/FRD data 
in their analysis and defined the IHV index 
as an average of the six absolute differences 
of the successive hourly values of the H component between 00 and 06 UT (19-01 LT).

%__________________Figure 3____________________________________________
%
\begin{figure*}
      \begin{center}
      \resizebox{\hsize}{!}{\includegraphics{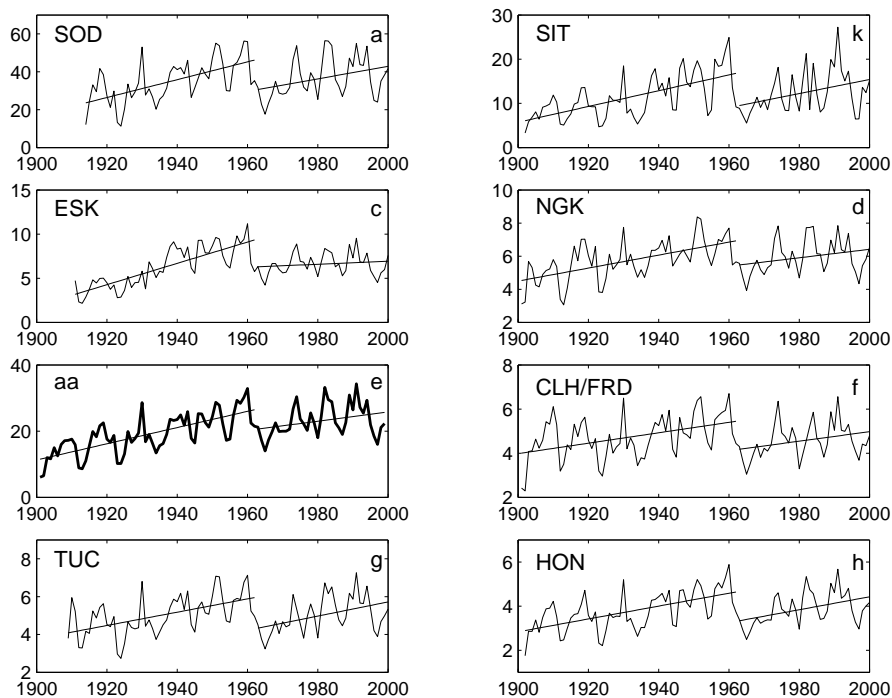}}
      \end{center}
      %\vskip -0.4cm
\caption{Yearly averages of the uncorrected IHV index (in nT) for the 7 stations
together with the aa index (panel e).
}
\label{Yearly_IHV}
%\vskip -0.3cm
\end{figure*}
%
%
%_______________________________________________________________________
%

We have calculated the IHV index (from the H component) for all the eight
stations of Table I to form seven long-term series of local geomagnetic activity.
(CLH and FRD IHV values were calculated separately and joined together 
in the same way as in Svalgaard et al., 2004).
Figure 3 depicts the seven series of yearly averaged IHV values 
and the aa index for comparison.
The absolute values of the IHV indices vary, as expected, greatly with the
magnetic latitude of the station so that the values 
at the highest SOD station are roughly
an order of magnitude larger than at the lowest HON station. 
Despite this difference, all IHV curves depict a clear solar cycle variation
which more closely follows the cyclicity of the aa index than of sunspots.
In fact, the IHV index seems to present a fairly good proxy of local 
geomagnetic activity, as suggested by Svalgaard et al. (2004).

Moreover, all the seven IHV series, as well as the aa index, 
depict the same qualitative long-term pattern during the last 100 years:
There is an increase of activity from the beginning of the 20th 
century until 1960, then a dramatic dropout in early 1960s, 
and (in most stations) weaker increasing activity therafter.
We have underlined this pattern in Figure 3 for each station by 
including the best fitting
line for the period until 1962 and another line for 1963-2000. 
Because of the (often overall) maximum in 1960 there is no uniform increase 
in geomagnetic activity during the last 100 years and
a two-line fit presents this step-like behaviour better than a one-line fit 
over the full time interval.

% TABLE II 
\begin{table}
%\begin{center}
%\begin{tabular}{|l|c|c|c|c|c|c|c|c|c|} \hline
\tiny
%\small
\begin{tabular}{lrrrrrrrrr} \hline
Station&\multicolumn{1}{c}{IHV} &\multicolumn{1}{c}{IHV} &\multicolumn{1}{c}{Rel.}ÊÊ &\multicolumn{1}{c}{IHV-q}Ê
&\multicolumn{1}{c}{IHV-q} &\multicolumn{1}{c}{Rel.}ÊÊ &\multicolumn{1}{c}{IHV-cor} &\multicolumn{1}{c}{IHV-cor}
&\multicolumn{1}{c}{Rel.} \\
 Ê&\multicolumn{1}{c}{start}Ê &\multicolumn{1}{c}{end}ÊÊ & \multicolumn{1}{c}{change} & \multicolumn{1}{c}{start}
& \multicolumn{1}{c}{end} & \multicolumn{1}{c}{change} & \multicolumn{1}{c}{start} & \multicolumn{1}{c}{end}ÊÊ
  & \multicolumn{1}{c}{change} \\
\hline
SOD & 28.65 & 40.00 & 40\% & 12.64 & 18.81 & 49\% & 16.01 & 21.20 & 32\% \\ %\hline
SIT &	8.59Ê & 13.89 & 62\%Ê &2.12ÊÊ &5.37ÊÊ&153\% & 6.48Ê & 8.52Ê & 31\%Ê \\ %\hline
ESK &	3.96Ê & 6.77Ê & 71\%Ê &1.51ÊÊ &1.24ÊÊ&-18\% & 2.45Ê & 5.52Ê & 125\%\\ %\hline
NGK & 5.11Ê & 6.19Ê & 21\%Ê &0.69Ê &0.82ÊÊ &19\%Ê &4.42Ê &5.37Ê &21\% \\ %\hline
CLH/FRD&4.47& 4.73Ê & 6\%Ê  &0.46Ê &0.47ÊÊ &2\%Ê &4.01Ê &4.26Ê &6\% \\ %\hline
TUC & 4.63Ê & 5.40Ê & 17\%Ê &1.08Ê &1.35ÊÊ &25\%Ê &3.55Ê &4.06Ê &14\% \\ %\hline
HON & 3.37Ê & 4.15Ê & 23\%Ê &0.87Ê &1.19ÊÊ &37\%Ê &2.50Ê &2.96Ê &18\% \\ \hline
\end{tabular}
\caption[]{Values of the IHV, IHV-q and IHV-cor indices in the beginning
(1901-1922) and at the end (1979-2000) of the last century.}
%\end{center}
\end{table}

However, the centennial increase is evident and in all the seven series
the IHV index attains a higher average level at the end of the century than 
in the beginning.
We have quantified this increase by calculating 
the average values of the IHV index during the last 
(1979-2000) and first (1901-1922) 22 years of the century.
(Because of different start years, the stations cover a 
slightly different fraction of the first 22 years). 
We have depicted these average levels as well as the 
implied percentual change (increase) of local geomagnetic activity in Table II.
This qualitatively uniform development in all the seven series proves that
geomagnetic activity has indeed increased during the last 100 years, as has
earlier been claimed on the basis of the aa index. 

Note, however, that the relative increase (column 4 of Table II) 
depicted by the seven series varies greatly, being higher at the high-latitude
stations (up to 70\% at ESK) and lower at the low to mid-latitude stations. 
While this difference can not yet be fully understood, it may  
indicate that the fraction of those disturbances 
in the solar wind that cause only moderate geomagnetic activity
(like substorms) observed mainly at high latitudes has increased during this time interval. 
Also, the CLH/FRD IHV series is exceptional in depicting by far the smallest
increase of about 6\% only, less than half of the second smallest increase at TUC
and a tenth of the increase at ESK and SIT. 
This exceptionally weak increase of local
geomagnetic activity at CLH/FRD  made Svalgaard et al. (2004) to  
erroneously doubt the increase of global geomagnetic activity
during the last 100 years.

Although we find that there is no doubt that global geomagnetic activity has
increased during the last 100 years, the exact amount of this increase is not
completely unambiguous. The corresponding relative increase in the aa index is
65\%, i.e.,  close to the percentual increase at SIT and ESK. 
However, a considerably lower increase is found at the mid-latitude
stations (NGK, CLH/FRD; about 50$^\circ$) that are closest in magnetic 
latitude to the aa index stations.
We can now make a conservative estimate on the centennial increase 
in global geomagnetic activity based on the results of Table II. 
Selecting one high-latitude (SIT), one mid-latitude (NGK) and one low-latitude
(HON) station from among those stations whose observations start either in 1901 or
1902, we obtain an average centennial increase of 35\%. 
(We neglect here CLH/FRD since, as mentioned by Svalgaard et al., 2004,
the values before 1915 may be overestimated).
This is also very close to the overall averaged increase of all stations of about 34\%.
Accordingly, we find that the centennial increase is
considerably lower, roughly by a factor of two, than depicted 
by the aa index.

%____________________SECTION_4______________

\section{Corrected IHV index}

As discussed above (see Figures 1 and 2), the range of daily curve 
varies strongly with solar activity (and geomagnetic activity at high latitudes).
Since the sunspot cycles depict a roughly similar long-term evolution 
(increase of cycle amplitudes until SC 19 in 1957; dropout and increase) 
as geomagnetic activity, it is clear that the daily range 
is also affected by this long-term change, as can also be seen in Figure 2.

Because of its very definition, the IHV index depends on the 
range of the daily curve during the 7 pre-midnight hours (19-01 LT).
Therefore, the long-term change of the daily range is expected to 
contribute to the long-term change of the IHV index.
Moreover, since geomagnetic activity is defined as 
a deviation from the quiet time
daily curve, the long-term variation in the daily range
has to be removed from the IHV index.
We have done this as follows.
We first calculated the yearly averaged daily curves for each station in order to
obtain a proxy for the quiet time daily variation in each year.
Then we calculated the yearly quiet-time IHV-q value from these 
smooth yearly curves in the same way as the IHV index is calculated for each day
from the raw data.

The averaged values of the IHV-q index at the start (1901-1922) and 
end (1979-2000) of the
100-year time interval (and the relative change) are included in Table II.
Note that the smallest IHV-q values are found for CLH/FRD, in agreement with the 
flatness of the daily curve during the IHV hours (see Figure 1).
Considerably larger IHV-q values are found in many other stations,
in particular in SOD where the daily curve is very steep in the
local pre-midnight hours (see Figure 1).
Note that in most stations the relative long-term change in IHV-q is 
positive and of the same order of magnitude as the relative change in IHV,
except at ESK where it is negative.

%__________________Figure 4____________________________________________
%
\begin{figure*}
      \begin{center}
      \resizebox{\hsize}{!}{\includegraphics{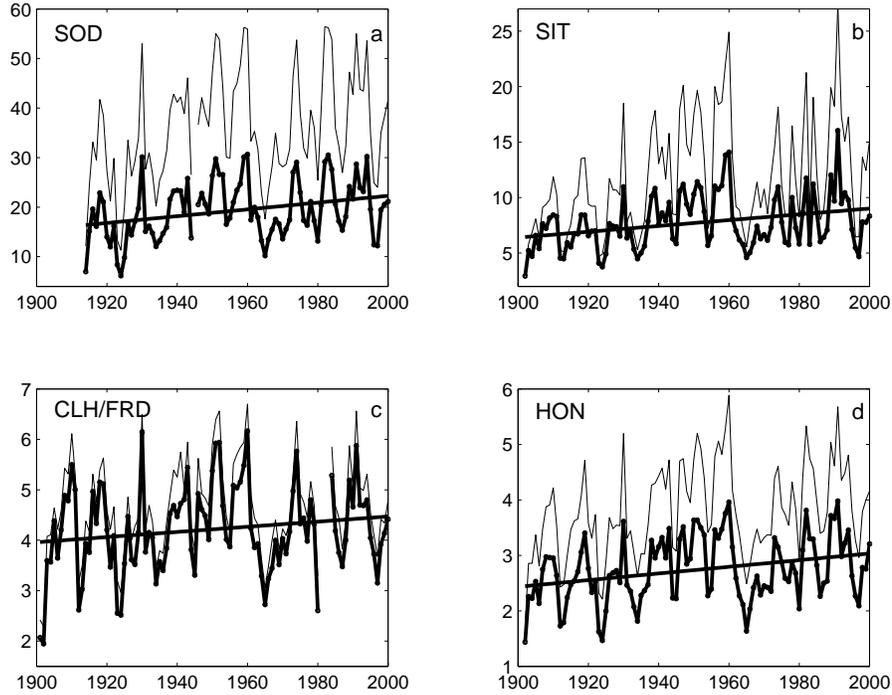}}
      \end{center}
      %\vskip -0.4cm
\caption{Yearly averages of the corrected (thick lines) and 
uncorrected (thin lines) IHV index for 4 stations.
Best fitting lines to IHV-cor are also included.
}
\label{Cor_uncor_IHV}
%\vskip -0.3cm
\end{figure*}
%
%_______________________________________________________________________
%

We have formed the corrected IHV-cor index by subtracting 
the yearly IHV-q values from the original daily IHV index. 
Figure 4 depicts the yearly averaged IHV-cor indices for a few stations
together with the original IHV index.
We have also depicted the best fitting lines to the IHV-cor series in Figure 4.
Note that the correction seems to be more important at high and low latitudes
while at mid-latitudes the correction is smaller.
The averaged values of the IHV-cor index at the start 
and end of the 100-year time interval (and the relative change) are included in
Table II. A faster long-term increase in IHV-q than in IHV leads 
to a smaller increase in the corrected IHV-cor index.
This is the situation in SOD, TUC, and HON where the centennial increase 
in IHV-cor is some 15-20\% smaller than in IHV.
Even a larger reduction is found for SIT where the increase in the corrected 
IHV-cor is nearly one half of the original IHV value.
In the other (interestingly, all mid-latitude) stations the reverse is true and there is 
a larger centennial increase in IHV-cor than in the original IHV.
The extreme is, because of the exceptional change of the daily curve, 
the ESK station where the IHVcor depicts some 60\% larger increase than IHV.
Thus, the net result of the daily curve correction is that the average 
centennial increase in the IHV-cor remains nearly the same, at about 35\%, 
as in the original IHV index.
However, if the centennial increase is calculated from the same
three longest stations as above (SIT, NGK, HON), a somewhat smaller value of about 24\%
is obtained for the IHV-cor.

Moreover, in order to further study the centennial increase
in the various stations we have normalized both the original and
the corrected IHV series at each station by their respective averages
and depicted the slopes of the best fitting lines in Table III.
This Table verifies the above results of the higher centennial increase 
at high latitudes and the different effect of the daily curve correction at
high and low latitudes vs. mid-latitudes.
Moreover, Table III also gives the average slope of the centennial increase
(and its standard deviation) at the seven stations.
Note that this average slope is roughly the same, about 0.0032,
both for the original and corrected IHV index.
This average slope corresponds to the average centennial increase
of about 38\%, in a very good agreement with the average increase
estimated above using the 22-year time intervals,
but remains clearly below the corresponding slope (about 0.0061)
for the aa index. 
These results further underline our main result
that  the centennial increase is indeed real but considerably 
 smaller than (roughly one half of) that depicted by the aa index.

% TABLE III 
\begin{table}
%\begin{center}
\tiny
%\small
%\begin{tabular}{lrrrrrrrrr} \hline
\begin{tabular}{lcccccccccc} \hline
Data& SOD& SIT& ESK& NGK& aa& CLH/FRD& TUC&	HON&	Mean& St.dev.\\
\hline
IHV/aa&	40& 55& 47& 24& 61&	11& 20& 26& 32& 16\\
IHV-cor&	35& 34& 77& 24&Ê  &	12& 19& 22& 32& 21\\
\hline
\end{tabular}
\caption[]{The slope (multiplied by 10$^4$) of the best fitting line 
for normalized IHV (original and corrected) values and aa index. 
Mean and standard deviation are obtained from IHV values only.}
%\end{center}
\end{table}

%____________________SECTION_5______________
\section{Conclusions}

We have reanalyzed geomagnetic activity during the last 100 years by  
calculating the recently suggested IHV index as a measure of 
local geomagnetic activity for 7 stations (or station pairs).
We find that local geomagnetic activity in all stations follows the same 
qualitative long-term pattern as depicted by the geomagnetic aa index: 
an increase from early 1900s to 1960, a dramatic dropout in 1960s
and a (mostly weaker) increase thereafter.
In all stations the average local geomagnetic activity 
during the last two decennia of the 20th century 
is stronger than during the first two decennia. 
This verifies the result based on the aa index that global
geomagnetic activity, and thereby, the open solar magnetic field 
has indeed increased during the last 100 years.

However, we find that the amount of centennial increase varies greatly from one
station to another. In particular, a very small increase is found in the CLH/FRD
series, which was erroneously interpreted in evidence of no centennial increase.
We find that the relative increase is higher at the 
high-latitude stations and lower at the low to mid-latitude stations. 
These differences may indicate that the fraction of 
 solar wind disturbances leading to only moderate geomagnetic activity
has increased during the studied time interval.
We have also shown that the IHV index needs to be corrected for the
long-term change of the daily curve, and we have calculated the corrected IHV
values.  In four out of seven stations this correction reduces
the estimated centennial increase of local geomagnetic activity.

Summarizing, our results give strong evidence that the centennial 
increase indeed took place during the last century.
However, the increase of about 65\% depicted in the aa index
is roughly twice larger than our more global estimate 
of about 24-35\% based on seven series of long-term observations.

\end{article}

\end{document}